\documentclass{NASspecial}

\usepackage[dvips]{graphicx}
\usepackage{amssymb,amsfonts,amsmath}

\def\lap{\lower.5ex\hbox{$\; \buildrel < \over \sim \;$}}
\def\gap{\lower.5ex\hbox{$\; \buildrel > \over \sim \;$}}

\url{www.pnas.org/cgi/doi/10.1073/pnas.0709640104}
\copyrightyear{2008}
\issuedate{Issue Date}
\volume{Volume}
\issuenumber{Issue Number}

\begin{document}

\title{Dark Matter}

\author{P. James E. Peebles\affil{1}{Princeton University, Princeton USA}}

\contributor{Submitted to Proceedings of the National Academy of Sciences of the United States of America}

\maketitle

\begin{article} 

\begin{abstract} 
The evidence for the dark matter of the hot big bang cosmology is about as good as it gets in natural science. The exploration of its nature is now led by direct and indirect detection experiments, to be complemented by  advances in the full range of cosmological tests, including judicious consideration of the rich phenomenology of galaxies. The results may confirm ideas about DM already under discussion. If we are lucky we also will be surprised once  again.
\end{abstract}


\dropcap{T}he case for the hypothetical nonbaryonic dark matter (DM) of the relativistic hot big bang $\Lambda$CDM cosmology, and its companion in the dark sector, Einstein's cosmological constant $\Lambda$ (or the near equivalent, dark energy, DE), rests in part on precision measurements, notably of the thermal cosmic microwave background radiation (the CMB). But equally important are the less precise measurements that look at the universe from many different sides and test for systematic errors in the measurements and inadequacies of the theory used to interpret the measurements.

The standard $\Lambda$CDM cosmology has six free parameters, which we may take to be the present cosmic mean mass densities (1) $\rho_{\rm b}$ and (2) $\rho_{\rm DM}$ in baryons and DM  (with $\Lambda$ fixed by the assumption of flat space sections), (3) the distance scale set by Hubble's constant $H_o$, (4) the tilt $n_s$ from scale-invariant primeval density fluctuations (taken to be adiabatic and Gaussian), (5) the amplitude of the primeval density fluctuations, and (6) the effective optical depth $\sigma$ for scattering of the CMB by intergalactic plasma after reionization started. It assumes textbook physics,  three neutrino families, no new relativistic species and no gravitational waves. Fitting $\Lambda$CDM to the precision CMB measurements  by the WMAP and PLANCK satellites\cite{WMAP,PLANCK} and ongoing ground- and balloon-based experiments\cite{ACT,SPT} requires models for foregrounds and instruments, and uses wide priors that prevent the fit from arriving at absurd situations such as a distance scale the astronomers would consider ridiculous (but I don't think the priors have much effect on the results). Under these conditions the CMB measurements alone constrain $\rho_{\rm b}$, $\rho_{\rm DM}$, and  $H_o$ to better than about $3\%$, and the optical depth $\sigma$ to about $15\%$, while the departure from scale-invariance measured by $n_s$ is detected at four standard deviations. The fit is impressive, ``nearly exhausting the information content of the temperature anisotropies''\cite{Planck15}. This is a reduction of a spectacular number of degrees of freedom in the CMB data to the six parameters plus well-motivated instrumental and cosmic noise estimates, with little room for anything else. 

The CMB certainly gives a good case for $\Lambda$CDM, but we should pause to consider that we are drawing conclusions about an immense universe from a modest data set. Consider also that this cosmology has in effect more than the advertised six free parameters. In the 1980s the community adopted the CDM cosmology over a considerable variety of alternatives then under discussion because CDM was seen to be promising. Since this was the choice of clever people we have to expect that the adopted model would enjoy some degree of success even if it were not on close to the right track. Also, we  felt free to adjust the cosmology to fit improving measurements, first by adding $\Lambda$, later by allowing departure from scale-invariance, in arriving at the present standard $\Lambda$CDM. If fitting the cosmology to the measurements had required replacing the constant $\Lambda$ with a function of time, or adding a soup\c{c}on of non-Gaussian or isocurvature departures from homogeneity, we would have done it. These choices of model and adjustments to be made or not made aided fitting theory to measurements; the effect is  hard to assess but certainly real. Of course, this is a normal and essential part of science. We make progress by ideas, and by measurements that inspire and attack ideas, a process that by repeated checks and successive approximations may reduce some ideas to theories that are so thoroughly tested as to qualify for entry in established canonical science. We have no prescription for quantifying the ``hidden free parameters'' in this process; the decision that predictions adequately outnumber parameters is a judgement, but one at which the scientific community has had a lot of practice. For examples of this judgement in action in cosmology, following \cite{WMAP,PLANCK}, let $\Lambda$CDM constrained by the CMB alone be a benchmark for assessment of all the other cosmological tests. The benchmark cosmic baryon mass density and baryon to DM mass ratio are 
\begin{align}
&\rho_{\rm b} = (4.14 \pm 0.05)\times 10^{-31}\hbox{ g cm}^{-3}, &\nonumber\\ 
&\rho_{\rm b}/\rho_{\rm DM} = 0.183\pm 0.005. &  \label{eq:benchmark}
\end{align}

The value of $\rho_{\rm b}$ was first seriously constrained\cite{GGST} in the 1970s, from estimates of the cosmic abundance of deuterium, which in the $\Lambda$CDM theory of nucleosynthesis of light elements in the hot big bang (BBNS) is sensitive to $\rho_{\rm b}$. The results argued against a cosmology with baryon mass density equal to the Einstein-de~Sitter value (in the relativistic model with no space curvature or $\Lambda$). The small value of $\rho_{\rm b}$ compared to the arguably elegant Einstein-de~Sitter case, and compared to estimates from dynamics of relative motions of galaxies, was one of the hints that led us to think of nonbaryonic DM. Now BBNS is a demanding  test of the CMB benchmark\cite{Iocco09}. The deuterium-to-hydrogen ratio is now measured from absorption lines in the spectrum of a background galaxy or quasar produced by the gas in a galaxy that happens to intersect the line of sight. In a recent study\cite{BBNS} the measured deuterium mass fraction fitted to BBNS requires 
\begin{equation}
\rho_{\rm b} = 4.19 \pm 0.17\hbox{ g cm}^{-3}.
\label{eq:BBNSprediction}
\end{equation}
This is less precise than the benchmark in equation~(\ref{eq:benchmark}), but the important point is that it is based on a very different set of considerations. Equation~(\ref{eq:BBNSprediction}) requires analysis of thermonuclear reactions as the $\Lambda$CDM universe expanded and cooled through temperature $T\sim 10^{10}\hbox{ K}$. Equation~(\ref{eq:benchmark}) requires analysis of effects of dynamics and diffusion on the distributions of DM, baryons, and radiation as the primeval plasma in the $\Lambda$CDM universe  cooled and combined largely to atomic hydrogen at $T\sim 4000\hbox{ K}$, and then beginning at $T\sim 100\hbox{ K}$ was ionized again, as matter and radiation moved through curved spacetime to the present epoch. That is, the fundamental theory is tested by application to two very different situations. The systematic error hazards are very different too. Stars can burn deuterium to helium, and chemistry can concentrate the deuterium in sites reached or not reached by absorption line observations, sources of systematic error. The deuterium abundance used in equation~(\ref{eq:BBNSprediction}) is based on observations of galaxies at redshift $z\sim 2$ to 3, seen when they were young because of the light travel time, and the heavy element abundances in these young galaxies are low, consistent with the idea that stars and chemistry have not had much time to alter the element abundances. The agreement of equations (\ref{eq:benchmark}) and (\ref{eq:BBNSprediction}) argues for this, but we do not know it for a fact, so how impressed you are by this test is a matter of judgement, to be aided by more tests. 

The second benchmark in equation (\ref{eq:benchmark}) may be compared to the baryon mass fractions in clusters of galaxies. The baryon mass estimate is the mass in plasma (based on its X-ray luminosity and spectrum and the inverse-Compton SZ effect on the CMB) added to the baryons in stars. The DM mass is estimated from the gravity needed to contain the plasma and the gravitational lensing of background galaxies. In the 1990s estimates of the cluster baryon mass fraction with $\rho_{\rm b}$ from BBNS indicated that $\rho_{\rm b} + \rho_{\rm DM}$ is well below Einstein-de Sitter. This would mean a relativistic cosmology requires space curvature or $\Lambda$, both of which were considered distasteful, as one sees in the subtitle of an early paper\cite{clustermassfraction}, {\it a challenge to cosmological orthodoxy}. Now $\Lambda$ is part of the orthodoxy. A recent estimate\cite{fb},
\begin{equation}
\rho_{\rm b}/\rho_{\rm DM} = 0.163\pm 0.032, \label{eq:clmassfraction}
\end{equation}
is within one standard deviation of equation~(\ref{eq:benchmark}). This is not going to become a precision measurement. For one thing, we cannot know whether numerical simulations capture all the process that change the cluster mass ratio from the cosmic mean. But it is a valuable measurement because the observations and their methods of analysis are so very different from the CMB. 

Since estimates of $\rho_{\rm b}$ and $\rho_{\rm DM}$ helped motivate $\Lambda$CDM the consistency of equations~(\ref{eq:benchmark}) to~(\ref{eq:clmassfraction}) is not a total surprise. But consider that when I proposed CDM and then $\Lambda$CDM in the 1980s\cite{pjep1,pjep2} I only aimed for CMB anisotropy below the limits we had then. It was not evident that the anisotropy would be detected, let alone measured in the detail for a benchmark. The  count of successful  $\Lambda$CDM predictions thus includes the fit to all the degrees of freedom in the CMB measurements, after discounting for the free choices. Also to be counted are the consistency of the CMB benchmark with applications of $\Lambda$CDM to  considerably improved measurements of the deuterium abundance in young galaxies and the baryon fraction in clusters of galaxies, and the list of other such demonstrations of consistency discussed in \cite{PlanckXVI}. An attempt at a still more complete list in Table 5.3 in\cite{FTBB} is five years old and largely out of date; it should be revisited. This network of checks is the case for DM and the rest of the standard cosmology. The precision tests are improving. I  urge the community to support work on the rest of the independent tests that are essential to the deepest possible examination of our universe. Maybe we will hit on anomalies that teach us something new; it has happened before.

It might appear that we are basing our theory of the universe on observations of a tiny part, the record in material in the narrow confines of our world path and in what happens to arrive on our incoming light cone. But consider that intergalactic plasma scatters CMB radiation into our line of sight from radiation that originated at a range of initial positions (with optical depth $\sigma\sim 0.09$ in $\Lambda$CDM). If this mix of initial positions presented us with a mix of different temperatures in the radiation received along a line of sight we would observe a nothermal CMB spectrum. The measured spectrum is impressively close to thermal (\cite{Kogut}, Fig. 1, from Alan Kogut, GSFC). That is, the initial temperature of the CMB, and the expansion that caused the temperature to evolve, have to have been quite close to homogeneous on scales smaller than the smoothing by scattering by intergalactic plasma. The CMB intensity is isotropic to a few parts in $10^5$, which allows a universe that is spherically symmetric about us with a radial temperature gradient small enough that scattering did not produce a noticeably nonthermal spectrum. But consider the observations of baryon acoustic oscillations (BAO), the ripples in the power spectrum of the matter distribution that accompany the peaks and valleys in the CMB temperature anisotropy spectrum. In $\Lambda$CDM they are are remnants of the acoustic oscillation of plasma and CMB that behaved as a fluid coupled by the short mean free path for Thomson scattering before recombination. BAO is measured in samples of galaxies at redshifts in the range $z\sim 0.3$ to 1 and in the Lyman~$\alpha$ forest at $z\sim 2$ to 3 (\cite{Eisenstein,BAO}). This probes what happened along  world paths of matter at a considerable range of distances from us. The $\Lambda$CDM fit to these samples is consistent with the CMB benchmark. This is evidence of large-scale homogeneity of initial conditions and of the $\Lambda$CDM physics that determined the BAO. You can add examples. In short, we have samples of the universe extending from our neighborhood to the Hubble length, and reaching from our incoming light cone back in time to the high redshift universe.  An analysis of the radial gradients in density and temperature that the cosmological tests would allow in a universe that is spherically symmetric about us would be interesting, but I see no reason to expect anything but a null result.

Galaxy phenomenology offers cosmological tests that could be particularly interesting because they probe smaller scales, though hazardous because strongly nonlinear evolution on the scale of galaxies complicates interpretations. An instructive example is Milgrom's\cite{Milgrom83} replacement of the hypothetical DM of $\Lambda$CDM by a hypothetical modification (MOND) of the Newton gravitational acceleration from an inverse square law $g\propto r^{-2}$ on small scales to $g\propto r^{-1}$ on large scales, in such a way that the circular velocity $v_c$ in the outer parts of a rotationally-supported galaxy scales with its baryonic mass $M_{\rm b}$ as 
\begin{equation}
M_{\rm b} \propto v_c^4.\label{eq:MOND}
\end{equation}
When Milgrom proposed this relation in 1983 it was known that the luminosity of a spiral galaxy scales  about as the fourth power of the circular velocity in its outer parts, and the luminosity of an elliptical scales about as the fourth power of its stellar velocity dispersion. I don't know whether Milgrom was aware of this, but it scarcely matters because in 1983 no one could have anticipated the tight consistency with equation~(\ref{eq:MOND}) found in late-type galaxies by replacing luminosity with the sum of the baryon masses in stars and atomic hydrogen and extending the test down to dwarf galaxies (as shown in Fig. 3 in \cite{MOND1}). MOND made a strikingly successful prediction. This is not an argument against $\Lambda$CDM, of course, unless it can be shown that the relation is improbable within $\Lambda$CDM. This is difficult to check because the outer parts of galaxies have complicated histories in $\Lambda$CDM (while in MOND there is  no DM and little of anything else in the outer parts of galaxies). Current numerical simulations of galaxy formation (e.g. \cite{Klypin}) suggest a reasonable but not yet convincing case for equation~(\ref{eq:MOND}) in $\Lambda$CDM. 

Though galaxy phenomenology is complicated we can make sensible judgements about  observations that may throw more light on the fundamental theory than the complexities of its expression. I like the example of pure disk galaxies, in which most of the stars move in streams in directions close to the plane of the disk, as in whirlpools and bars\cite{Kormendy}. This is different from galaxies that have a rotationally-supported disk and centered on it a classical bulge of stars supported largely by random motions, as in an elliptical galaxy. The nearby spiral M81 has a classical bulge. Our Milky Way seems to be a pure disk galaxy, the stars even near the center streaming in near rotational support\cite{Rich}. The Milky Way also has an extended stellar halo supported by near random motions, but its luminosity is only a few percent of the total. The classical bulge of M81 seems to be a natural product of evolution in $\Lambda$CDM, which predicts the growth of galaxies by mergers and accretion at redshifts $1<z<3$, at a time when the global star formation rate was high. The stars that formed then in the bits and pieces that were flowing into a growing galaxy would end up in a stellar halo or classical bulge. This is seen in numerical simulations; an example\cite{Madau} is an elegant approximation to M81, with a distribution of streaming and near random velocities characteristic of a disk with a classical bulge (as reflected in the two peaks in Fig.~6 in \cite{Madau}). Thin disk galaxies are the puzzle. Despite all the complications of baryon physics we can be reasonably sure that the stars that formed in the bits and pieces flowing into a growing galaxy would not end up streaming in the disk; that requires dissipative settling prior to star formation. Can $\Lambda$CDM explain how material streaming together to form a pure disk galaxy ``knew'' that when the star formation rate was high star formation had to be almost entirely confined to the one fragment that is going to grow into the present-day disk? I offer this puzzle as a counterexample to the proposition that galaxies are too complicated to add to the probes of fundamental theory. We certainly will learn something of value from close considerations of thin disk galaxies, the scaling law in equation~(\ref{eq:MOND}), and many other issues of galaxy phenomenology. Maybe we will learn something about DM. 

It is good to question authority, and fair to ask\cite{MOND2} whether there is DM-free cosmology that fits the tests as well as $\Lambda$CDM, perhaps one built on the MOND prediction of equation~(\ref{eq:MOND}).  We cannot disprove it. For that matter, we cannot disprove existence of an alternative physics that does as well as what is in the textbooks, we can only note that it looks wildly unlikely. The broad suite of successful $\Lambda$CDM predictions on scales larger than those characteristic of galaxies makes a DM-free alternative seem exceedingly unlikely too. But it is easy to imagine that $\Lambda$CDM is not the final theory for cosmology at redshifts $z\lap 10^{10}$, only the simplest approximation we can get away with at the present level of evidence.  A still better cosmology might have modified gravity, maybe dynamic DE, maybe more interesting DM. The many ideas about the last include some mixture of cold and warm components (we already have a hot component, neutrinos), self-interacting or annihilating, decaying with or without emission into the visible or dark sectors, charged or strongly interacting, in condensates or black holes, and with fanciful names such as axions, WIMPs, WIMPZILLAs, fuzzy particles, and Q-balls, motivated by ideas that are interesting and may prove to be relevant. Of course, the essential complement  to relevant ideas is relevant evidence, from the direct and indirect DM searches, and I expect from anomalies in the fit of $\Lambda$CDM to the full variety of cosmological tests. 

Cosmology has enjoyed a wonderfully productive growth spurt in the last quarter century, yielding the network of tests that place DM and $\Lambda$ (or DE) in the canon of persuasively established physics. This great advance leaves us with a great opportunity, to explore the dark sector. There will be more growth spurts, and a clearer picture of the dark matter.

\end{article}

\end{document}